\documentclass[a4paper, 11pt]{article}

\usepackage{a4wide}
\usepackage{amsmath}
\usepackage{latexsym}
\usepackage{graphicx}
\usepackage{color}
\usepackage{hyperref}
\usepackage{verbatim}

\title{A stochastic evolutionary model for capturing human dynamics}

\author{Trevor Fenner, Mark Levene, and George Loizou \\
Department of Computer Science and Information Systems \\
Birkbeck, University of London \\
London WC1E 7HX, U.K. \\
\{trevor,mark,george\}@dcs.bbk.ac.uk}

\date{}

\begin{document}








\maketitle

\begin{abstract}

The recent interest in human dynamics has led researchers to investigate the stochastic processes that explain human behaviour in various contexts. Here we propose a generative model to capture the dynamics of survival analysis, traditionally employed in clinical trials and reliability analysis in engineering.
We derive a general solution for the model in the form of a product, and then a continuous approximation to the solution
via the renewal equation describing age-structured population dynamics. This enables us to model a wide range of survival distributions,
according to the choice of the mortality distribution.
We provide empirical evidence for the validity of the model from a longitudinal data set of popular search engine queries over 114 months, showing that the survival function of these queries is closely matched by the solution for our model with power-law mortality.

\end{abstract}


\section{Introduction}

Recent interest in complex systems, such as social networks, the world-wide-web, email networks and mobile phone networks \cite{BARA07},
has led researchers to investigate the processes that could explain the dynamics of human behaviour within these networks.
Barab\'asi \cite{BARA05} suggested that the bursty nature of human behaviour, for example, when measuring the inter-event response time of email communication, is a result of a decision-based queuing process. In particular, humans tend to prioritise actions, for example, when deciding which email to respond to, and therefore a priority queue model was proposed in \cite{BARA05}, leading to a heavy-tailed power-law distribution of inter-event times.
The availability of large data sets, such as mobile phone records, has widened the applicability of human dynamics investigation, for example, in an attempt by Schneider et al. \cite{SCHN13} to uncover the characteristics of daily mobility patterns.
Human dynamics is not limited to the study of behaviour within communication networks, as can be seen, for example, by a recent proposal of Mitnitski et al. \cite{MITN13}, who apply a simple stochastic queueing model to the complex phenomenon of aging in order to illustrate how health deficits accumulate with age.

\smallskip

Survival analysis \cite{KLEI12} provides statistical methods to estimate the time until an event occurs, known as the {\em survival time}. Typically, an event in a survival model is referred to as a {\em failure}, as it often has negative connotations, such as mortality or the contraction of a disease, although it could, in principle, also be positive, such as the time to return to work or to recover from a disease. In the context of email communication mentioned above, an event might be a reply  to an email. Traditional applications of survival analysis are in clinical trials \cite{FLEM00}, and in understanding the mechanisms in biological ageing \cite{GAVR01}. The methods used in survival analysis overlap with those used in engineering for reliability life data analysis \cite{OCON12}.
Reliability engineering has many applications, for example, in manufacturing processes, and in software design and testing.
However, one can envisage that survival analysis would find application in newer human dynamics scenarios in complex systems, such as those arising in social and communication networks \cite{BARA05,CAND08,MATH13}.

\smallskip

Of particular interest to us has been the formulation of a {\em generative model} in the form of a stochastic process by which a complex system evolves and gives rise to a power law or other distribution \cite{FENN05,FENN12}.
This type of research builds on the early work of Simon \cite{SIMO55}, and the more recent work of Barab\'asi's group \cite{ALBE01} and other researchers \cite{BORN07a}. In the context of human dynamics, the priority queue model \cite{BARA05} mentioned above is a generative model characterised by a heavy-tailed distribution.

In the bigger picture, one can view the goal of such research as being similar to that of {\em social mechanisms} \cite{HEDS98}, which looks into the processes, or mechanisms, that can explain observed social phenomena. Using an example given in \cite{SCHE98a}, the growth in the sales of a book can be explained by the well-known logistic growth model \cite{TSOU02}, and more recently we have shown that the process of conference registration with an early bird deadline can be modelled by bi-logistic growth \cite{FENN13}.
Such research can also be grounded within the field of sociophysics \cite{GALA08,CAST09}, which applies methods from physics to explain social phenomena such as opinion dynamics. Critical phenomena are important here, where a transition to global behaviour emerges from the interactions of many individuals.
The individuals may be neurons as in \cite{LOVE12}, where criticality emerges as neuron cooperation,
or people's decisions as in \cite{PAN10}, where the popularity of movies emerges as collective choice behaviour.

\smallskip

In \cite{FENN14} we proposed a simple generative model to capture the essential dynamics of survival analysis applications. For this purpose, we make use of an urn-based stochastic model, where the {\em actors} are called {\em balls}, and a ball being present in $urn_i$, the $i$th urn, indicates that the actor represented by the ball has so far survived for $i$ time steps. An actor could, for example, be a subject in a clinical trial, an email that has not yet been replied to, or an ongoing phone call. As a simplification, we assume that time is discrete and that, at any given time, one ball may join the system with a fixed {\em birth} probability. Alternatively, with a fixed probability, an existing ball in the system may be chosen uniformly at random and discarded, i.e. a {\em mortality} (or death) event occurs. It is evident that, at any given time, say $t$, we may have at most one ball in $urn_i$, for all $i \le t$.

\smallskip

The main result in \cite{FENN14} was to derive a power-law distribution for the probability that, after $t$ steps of the stochastic process outlined above, there is a surviving ball in $urn_i$. Thus, in our model, the {\em survival function} \cite{KLEI12}, which gives the probability that an actor (in our model, a ball) survives for more than a given time, can be approximated by a power-law distribution. It is interesting to observe that the resulting distribution has two parameters, $i$ and $t$, as in \cite{FENN12}, whereas most previously studied generative stochastic models \cite{ALBE01,NEWM05}, including those in our previous work, for example \cite{FENN05}, result in steady state distributions that are asymptotic in $t$ to a distribution with a single parameter $i$.

\smallskip

Here we relax the stipulation that balls are discarded according to a uniform distribution and allow the probability of mortality to take a general form.
This will enable us to derive a wide range of distributions from the generative model, and of particular interest is the case when mortality follows a power-law distribution. Our model can be described by a difference equation that has an explicit solution in the form of a product.
The difference equation can be approximated by a partial differential equation that coincides with the renewal equation from age-structured population dynamics with a constant birth rate \cite{CHAR94,LI08}. More specifically, we show that, by choosing the appropriate mortality distribution, the survival distribution of actors will follow an exponential, power-law or Weibull distribution. The Weibull distribution \cite{RINN09} is of particular interest due to its prevalence in modelling {\em life data} (also known as survival data, time to event data, or time to failure data) \cite{KLEI12,OCON12}. It also has application in the instance theory of automaticity \cite{COLO95}, where it is argued that the retrieval time of traces from memory follow a Weibull distribution, and more recently as a model for Internet response times \cite{CHES06}, where it is shown to be a better fit to the data than a power-law distribution.
We provide empirical evidence for the validity of the model from a longitudinal data set of popular search engine queries over 114 months, showing that the survival function of these queries closely follows the distribution generated by our model.

\smallskip

The rest of the paper is organised as follows. In Section~\ref{sec:urn}, we present our stochastic urn-based model that provides us with a mechanism to model the essential dynamics of survival models. We then derive a difference equation to describe the process and obtain its solution in the form of a product.
In Section~\ref{sec:pop}, we provide a continuous approximation to the model, which has a solution in the form of an integral.
In Section~\ref{sec:mortality}, we derive the resulting distributions in the continuous case for several mortality distributions; in particular, when the mortality distribution is a power law, we obtain a Weibull distribution.
In Section~\ref{sec:survival}, we apply our generative model to the survival of popular search engine queries posted on Google Trends (\url{www.google.com/trends/hottrends}). Finally, in Section~\ref{sec:conc}, we give our concluding remarks.

\section{An Evolutionary Urn Transfer Model}
\label{sec:urn}

In this section we formalise a stochastic urn model that allows us to model the dynamic aspects of a complex system, and then present the difference equations that describe the model. This model is a direct extension of the one introduced in \cite{FENN14}, in that it caters for an arbitrary mortality distribution.

\smallskip

We assume a countable number of urns, $urn_0, urn_1, \ldots \ $, where a {\em ball} (or {\em actor}) in $urn_i$ is said to be of {\em age} $i$.
Initially, all the urns are empty.
We define a stochastic process \cite{ROSS96} where, at any time $t$, $t \ge 1$, a new ball may be {\em born} by inserting it into $urn_0$, and an existing ball of age $i$ can {\em die} by being discarded from $urn_i$, for all $i > 0$.

For a given age $i$ and time $t$, we let $\mu(i,t)$ be the probability that a ball in $urn_i$ dies at time $t$; $\mu(i,t)$ is known as the {\em mortality distribution}. We always require that $\mu(0,t)=0$ for all $t$. Finally, $urn_i$ is empty when $i > t$.

\smallskip

At time $t$, the stochastic process then proceeds as follows in order to obtain the configuration at time $t+1$, where $t \ge 1$.
\renewcommand{\labelenumi}{(\roman{enumi})}
\begin{enumerate}
\item For each $i$, $0 \le i \le t$, if $urn_i$ is non-empty then, with probability $\mu(i,t)$, the ball in $urn_i$ is discarded.

\item Next, the ages of all balls remaining in the system are incremented by $1$, i.e. a ball in $urn_i$ is moved to $urn_{i+1}$, for each $i$.

\item Finally, with probability $p$, where $0 < p < 1$, a birth occurs, i.e. a ball is inserted into $urn_0$.
\end{enumerate}
\smallskip


\medskip

We now let $F(i,t) \ge 0$ be a discrete function denoting the probability that there is a ball in $urn_i$ at time $t$. Initially, we set
$F(0,0) = p$ and $F(i,0) = 0$ for all $i > 0$.

\smallskip

The dynamics of the model can be captured by the following two equations:
\begin{equation}\label{eq:init}
F(0,t) = p \ \ {\rm for} \ \ t \ge 0,
\end{equation}
and
\begin{equation}\label{eq:diff}
F(i+1,t+1) = F(i,t) - \mu(i,t) F(i,t) \ \ {\rm for} \ \ 0 \le i \le t.
\end{equation}
\smallskip

We can expand (\ref{eq:diff}) to obtain
\begin{equation}\label{eq:expand}
F(i+1,t+1) = p \prod_{j=1}^{i} \left( 1 - \mu(j,t-i+j) \right).
\end{equation}

\section{A Stochastic Model Based on Population Dynamics}
\label{sec:pop}

In this section, we present an approximate solution to the difference equation (\ref{eq:diff}), by approximating it by a first-order hyperbolic partial differential equation \cite{LAX06}. This equation is the same as that used in age-structured models of population dynamics \cite{CHAR94}.
We note that, in our model, the birth rate $p$ is constant, rather than the more complex dynamics where the birth rate is determined by a distribution that depends on the ages of individuals and possibly on the time $t$.

\smallskip

We first rewrite (\ref{eq:diff}) in the form:
\begin{equation}\label{eq:diff-partial}
F(i+1,t+1) - F(i+1,t) + F(i+1,t) - F(i,t) + \mu(i,t) F(i,t) = 0,
\end{equation}
and then approximate the discrete function $F(i,t)$ by a continuous function $f(i,t)$, with $\mu(i,t)$ now being a continuous density function. We then approximate
\begin{displaymath}
F(i+1,t+1) - F(i+1,t) \ \ {\rm by} \ \ \frac{\partial f(i,t)}{\partial t}
\end{displaymath}
and
\begin{displaymath}
F(i+1,t) - F(i,t) \ \ {\rm by} \ \ \frac{\partial f(i,t)}{\partial i}.
\end{displaymath}
\smallskip

From (\ref{eq:diff-partial}) we thus derive the first-order hyperbolic partial differential equation,
\begin{equation}\label{eq:partial}
\frac{\partial f(i,t)}{\partial t} + \frac{\partial f(i,t)}{\partial i} + \mu(i,t) f(i,t) = 0.
\end{equation}
\smallskip

Note that, by (\ref{eq:init}), $f(0,t) = p$ for all $t$.

\smallskip

Equation (\ref{eq:partial}) is the well-known {\em transport equation} in fluid dynamics \cite{LAX06}, and
the {\em renewal equation} in population dynamics \cite{PILA91,CHAR94,LI08}.
Following Equation $1.22$ in \cite{CHAR94}, the solution of (\ref{eq:partial}), when $i \le t$, is given by
\begin{equation}\label{eq:renewal}
f(i,t) = f(0, t-i) \exp \left( - \int_{0}^{i} \mu \left( i-s, t-s \right) ds \right).
\end{equation}

\section{Choosing the mortality distribution}
\label{sec:mortality}

In this section, we make use of the generality of our model by investigating four special cases of the mortality distribution. We note that
$\mu(i,t)$ can be viewed as a {\em discrete hazard function} \cite{BRAC03}, denoting the conditional probability that an actor of age $i$ fails at time $t$, given that this actor did not fail previously.

\smallskip

In Subsection~\ref{subsec:const}, we look at the case when the mortality distribution is constant.
In Subsection~\ref{subsec:uni}, we look at the case when the mortality distribution is independent of age.
In Subsection~\ref{subsec:pref}, we look at the case when the mortality distribution is preferential.
Finally, in Subsection~\ref{subsec:power}, we look at the case when the mortality distribution is a power law in the age of the actor.

\subsection{Constant mortality}
\label{subsec:const}

In the simplest case, we let
\begin{equation}\label{eq:const1}
\mu(i,t) = C,
\end{equation}
for $i  >0$ and some positive constant $C$.

\smallskip

Substituting this into (\ref{eq:renewal}) and using the boundary condition (\ref{eq:init}), we obtain
\begin{displaymath}
f(i,t) = p \exp\left(- C i \right),
\end{displaymath}
which is the survival function of the exponential distribution, with rate parameter $C$.

\subsection{Age-independent mortality}
\label{subsec:uni}

Let
\begin{equation}\label{eq:uni}
\mu(i,t) = \mu(t) = \frac{\kappa}{t},
\end{equation}
for $i > 0$ and some positive constant $\kappa$.
In this case mortality does not depend on $i$, the age of the actor, so all actors who are alive at time $t$ are equally likely to die.

\smallskip

Substituting (\ref{eq:uni}) into (\ref{eq:renewal}), we obtain
\begin{align*}
f(i,t) &= p \exp\left(- \int_{0}^{i} \frac{\kappa}{t-s} \right) ds \\
       &= p \exp\left(\kappa \ln \left( \frac{t-i}{t} \right) \right) \\
       &= p \left( 1 - \frac{i}{t} \right)^{\kappa},
\end{align*}
which is a power-law distribution in $(t-i)/t$, as was derived from first principles in \cite{FENN14}.

\subsection{Preferential mortality}
\label{subsec:pref}

Let
\begin{equation}\label{eq:pre}
\mu(i,t) = \frac{\kappa i}{t^2},
\end{equation}
where $\kappa$ is a constant. In this case mortality is a function of both $i$ and $t$, where at a given time instant an older actor is more likely to die than a younger one. It is interesting to note that preferential mortality, being proportional to the age $i$, has some resemblance to the preferential attachment rule in evolving networks \cite{ALBE01}, where the probability that a node (or actor) gains a new link is proportional to the number of links it already has.

\smallskip

Substituting (\ref{eq:pre}) into (\ref{eq:renewal}) we obtain
\begin{align*}
f(i,t) &= p \exp\left(- \int_{0}^{i} \frac{\kappa (i-s)}{(t-s)^2} \right) ds \\
       &= p \exp\left(-  \kappa \left( \ln\left(\frac{t}{t-i}\right) - \frac{i}{t} \right) \right) \\
       &= p \exp\left( \frac{\kappa i}{t} \right) \left( 1 - \frac{i}{t} \right)^{\kappa},
\end{align*}
which is a power-law distribution with an exponential correction.

\subsection{Power-law mortality}
\label{subsec:power}

Let
\begin{equation}\label{eq:power}
\mu(i,t) = \mu(i) = \lambda (1+ \rho) \ i^\rho,
\end{equation}
for $i > 0$, and some {\em shape} parameter $\rho$, $-1 \le \rho \le 0$, and {\em scale} parameter $\lambda >0$.
We note that, in this case, mortality is time invariant.

\smallskip

Substituting (\ref{eq:power}) into (\ref{eq:renewal}) we obtain
\begin{align}\label{eq:weibull}
f(i,t) &= p \exp\left(- \int_{0}^{i} \lambda (1+ \rho) \ \left(i-s\right)^\rho \right) ds \nonumber \\
       &= p \exp\left(- \lambda \ i^{1+ \rho} \right),
\end{align}
which, when divided by $p$, is the survival function of the Weibull distribution \cite{RINN09}, also known as the stretched exponential function \cite{LAHE98}.
(We note that our definition of the scale parameter is slightly different from that given in \cite{RINN09}.)
The Weibull distribution is widely used in survival models \cite{KLEI12} and reliability engineering \cite{OCON12},
and it is therefore important to be able to model it.

\smallskip

Inspecting (\ref{eq:power}), we note that, in our model, the probability of mortality decreases with age.
The survival probability $f(i,t)$ also decreases with age, and decreases faster for larger $\rho$. We observe that this degenerates to the constant mortality case when $\rho=0$, and that $f(i,t)$ approaches a constant as $\rho$ gets close to $-1$.

\smallskip

Substituting (\ref{eq:power}) into (\ref{eq:expand}), with $k \ge 0$, we obtain
\begin{equation}\label{eq:solve-power}
\frac{F(i+1,t+1)}{F(k,t+1)} = \prod_{j=k}^{i} \left( 1 - \lambda (1+ \rho) j^\rho \right).
\end{equation}
\smallskip

Taking logarithms  in (\ref{eq:solve-power}) and using  the approximation $\ln(1+x) \approx x$, which holds for small $x$, we obtain
\begin{equation}\label{eq:log-F}
\ln \frac{F(i+1,t+1)}{F(k,t+1)} \approx - \sum_{j=k}^i \lambda (1+ \rho) j^\rho.
\end{equation}
\smallskip

Using the first two correction terms of the Euler-Maclaurin summation formula \cite{APOS99,LAMP01} to approximate the right-hand side of (\ref{eq:log-F}) we obtain
\begin{align}\label{eq:eulerk}
\sum_{j=k}^i \lambda (1+ \rho) j^\rho  & \approx \int_k^i \lambda (1+ \rho) x^\rho dx +
\frac{\lambda (1+ \rho) \left( i^\rho + k^\rho \right)}{2} + \frac{\lambda (1+ \rho) \rho \left( i ^{\rho-1} - k^{\rho-1} \right)}{12} \nonumber \\
& = \lambda \left( i^{1+ \rho} - k^{1 + \rho} \right) +
\frac{\lambda (1+ \rho) \left( i^\rho + k^\rho \right)}{2} + \frac{\lambda (1+ \rho) \rho \left( i ^{\rho-1} - k^{\rho-1} \right)}{12}.
\end{align}
\smallskip

Substituting (\ref{eq:eulerk}) into (\ref{eq:log-F}) and  rearranging, we obtain
\begin{equation}\label{eq:weibullk}
- \lambda i^{1+ \rho} \approx \ln \left( \frac{F(i+1,t+1}{F(k,t+1)} \right) - \lambda k^{1 + \rho} +
\frac{\lambda (1+ \rho) \left( i^\rho + k^\rho \right)}{2} + \frac{\lambda (1+ \rho) \rho \left( i ^{\rho-1} - k^{\rho-1} \right)}{12}.
\end{equation}

\section{Application to the Survival of Popular Search Engine Queries}
\label{sec:survival}



As mentioned in the introduction, survival analysis \cite{KLEI12}, dealing with the analysis of {\em time-to-event} data, is well
established within the statistics community, and has many applications in disparate fields. In the context of human dynamics, survival analysis has recently been applied to large data sets. These include the analysis of phone call durations \cite{MELO10}, the investigation of how long Wikipedia editors remain active \cite{ZHAN12},
and the analysis of completion rates for students using intelligent tutoring systems \cite{EAGL14}.

\smallskip

In our model, the objects being monitored are represented by balls and they are considered to have survived for as long as they remain in the system. A death event is modelled by discarding a ball, and a birth event is modelled by inserting a new ball into the first urn.
Our stochastic model has three input parameters: the birth probability $p$, the mortality distribution $\mu(i,t)$, and the time $t$ at which the system is observed. Given these parameters, the survival probability of a ball in $urn_i$ at time $t$, where $i \le t$, is approximated by $f(i,t)$ as given by (\ref{eq:renewal}), which depends on the form of the mortality distribution $\mu(i,t)$. In other words, given $t$, $f(i,t)$ is the probability that a new ball enters the system at time $t-i$ and survives for at least $i$ steps before it is discarded. In our empirical analysis below, we will make use of power-law mortality, which leads to the Weibull distribution, as in (\ref{eq:weibull}).

\smallskip

In survival analysis, we are often interested in the {\em survival function} $S(\theta)$ \cite{KLEI12}, which represents the probability that a patient in a given study survives for longer than a specified time $\theta$. The survival function is usually estimated via a step function by computing the probability that a patient survives until time $\theta$, for $\theta = 1,2,\ldots, t$. This step function is known as the {\em Kaplan-Meier estimator} \cite{KAPL58,KLEI12}. By comparing  (\ref{eq:solve-power}), or the more general (\ref{eq:expand}), with the Kaplan-Meier estimator for the survival function \cite[equation~(2b)]{KAPL58}, the latter is seen to be analogous to $F(i,t)$ for an actor that was born at time $t-i$; more specifically, $S(i) \approx F(i,t)/p$.

\smallskip

We note that, although in theory the survival function $S(\theta)$ does not depend on the length $t$ of the trial, in practice the Kaplan-Meier estimate will be more accurate for longer trials. On the other hand, this estimate is more accurate when most of the patients are still present in the study, since, when there are only a few patients left, the estimate may be inaccurate \cite{RICH10}.

\medskip

The Kaplan-Meier estimator also takes into account {\em censored} data \cite{KLEI12}, when, for example, a patient drops out before the end of the study period. Although our evolutionary urn transfer model does not explicitly include censoring, it could be incorporated by allowing the possibility that when a ball is discarded it may be counted as either a death or censoring event. We further note that, while in traditional survival models patients join a study in batches, in our model individual balls continue to join the system with a fixed probability. Our model could be generalised to allow several balls to join the system at any given time, and also by letting the arrival probability $p$ depend on $t$; we leave consideration of such enhancements for future work.

\medskip

As a proof of concept for the model with power-law mortality, we analyse the survival of queries in the top-10 Google Trends ``hot searches''  (\url{www.google.com/trends/hottrends}). Data was collected monthly for the top-10 ``hot searches'' over 114 months from January 2004 until June 2013,
for six categories (together with their subcategories in each case): Business \& Politics (or simply Business), Entertainment, Nature \& Science (or simply Science), Shopping \& Fashion (or simply Shopping), Sports, and Travel \& Leisure (or simply Travel). The number of distinct queries per category over the period is shown in Table~\ref{table:google}. It is apparent from this statistic that the top-10 queries from Shopping change the least, while those from Entertainment change the most.

\begin{table}[ht]
\begin{center}
\begin{tabular}{|l|c|}\hline
Data set       & Number of queries \\ \hline \hline
Business       & 318 \\ \hline
Entertainment  & 1672 \\ \hline
Science        & 150 \\ \hline
Shopping       & 107 \\ \hline
Sports         & 774 \\ \hline
Travel         & 342 \\ \hline \hline
All Categories & 3363 \\ \hline
\end{tabular}
\end{center}
\caption{\label{table:google} Number of top-10 queries collected from Google Trends.}
\end{table}
\smallskip

In this data set, the balls are top-10 queries, a death event occurs when a query leaves the top-10 in a given month, and a birth event occurs when a new query joins the top-10 in a given month (note that time is discrete and is measured in months).
As we are examining power-law mortality, queries leave the top-10 according to (\ref{eq:power}).
Thus, when $\rho$ is non-zero, the probability of mortality decreases with age.
As a result, a substantial number of queries are popular for a short duration, while the popularity of others lasts for much longer;
we note that a similar observation, in the context of the popularity of movies, was made in \cite{PAN10}.
Returning to our theme of human dynamics \cite{CAST09}, our stochastic urn model with power-law mortality contributes to a
better understanding of collective phenomena, such as popularity, and how such global behaviour emerges from the decisions of individuals.

\medskip

We first outline the methodology we have used to validate and evaluate our model, assuming power-law mortality as in (\ref{eq:power}).
We then give further details, before discussing and analysing the results.

\renewcommand{\labelenumi}{(\Roman{enumi})}
\begin{enumerate}
\item First, to obtain estimates of $\lambda$ and $\rho$, we perform least-squares curve fitting to the values of the product on the right-hand side of (\ref{eq:solve-power}) for $i = 1,2,\ldots,114$, with $k=0$, using the Kaplan-Meier estimates computed from the raw data.
\item We use $\lambda$ and $\rho$ from (I) to compute, for each $i$, the product on the right-hand side of (\ref{eq:solve-power}), with $k=0$; we call this the {\em product data}. We then repeat the least-squares curve fitting using these values as a quick check that the $\lambda$ and $\rho$ obtained are consistent with those from (I).
\item Next we run simulations using the parameters $\lambda$ and $\rho$ from (I), and $p=0.9$. Using the averaged values of $F(i,t)$ from the simulations, we again repeat the least-squares curve fitting, with $k=0$, to obtain new values for $\lambda$ and $\rho$; these are compared to those from (I) for consistency.
\item We compute the $D$ values from a Kolmogorov-Smirnov test to check whether the Kaplan-Meier estimates, the product data and the averaged simulation data are likely to have all come from the same distribution.
\item Lastly, we adjust the averaged values of $F(i,t)$ from the simulation using the Euler-Maclaurin correction, as on the right-hand side of (\ref{eq:weibullk}), with $k=10$ and the values of $\lambda$ and $\rho$ obtained from step (III). We then use nonlinear regression to fit a Weibull distribution to the exponential of the adjusted estimates in (\ref{eq:weibullk}). We compare the $\lambda$ and $\rho$ from this Weibull fit to those from (III), in order to check the plausibility of using the Weibull distribution as a continuous approximation to our model.
\end{enumerate}
\smallskip

We first computed the Kaplan-Meier estimates from the raw data sets for the individual six categories and for their aggregation (All Categories). Recalling that the survival function $S(i) \approx F(i,t)/p$, following (I), using Matlab we then obtained estimates for $\lambda$ and $\rho$ by non-linear least-squares regression of $S(i)$ on $i$ for fitting  function (\ref{eq:solve-power}), for $i = 1,2,\ldots,114$, with $k=0$.
The estimated parameters $\lambda$ and $\rho$ are shown in the rows of Table~\ref{table:lsq-km}, together with the coefficient of determination $R^2$ \cite{MOTU95}; these show a very good fit for all of the categories. We observe that the $R^2$ values for Science and Shopping are somewhat worse than the others, which could be attributed to the fact the number of queries in these categories is significantly smaller than the others, as can be seen from Table~\ref{table:google}.
Nonlinear least-squares curve fitting to the product data obtained using (\ref{eq:solve-power}), as in (II), yields perfect fits, as expected.

\begin{table}[ht]
\begin{center}
\begin{tabular}{|l|c|c|c|}\hline
Data set       & $\lambda$ & $\rho$  & $R^2$\\ \hline \hline
Business       & 1.5576    & -0.8321 & 0.9818 \\ \hline
Entertainment  & 1.7877    & -0.7453 & 0.9957 \\ \hline
Science        & 0.7575    & -0.8318 & 0.9303\\ \hline
Shopping       & 0.4736    & -0.7608 & 0.9240 \\ \hline
Sports         & 3.1986    & -0.8666 & 0.9967 \\ \hline
Travel         & 3.5141    & -0.9291 & 0.9760 \\ \hline
All Categories & 4.5262    & -0.9134 & 0.9955 \\ \hline
\end{tabular}
\end{center}
\caption{\label{table:lsq-km} Nonlinear least-squares regression with fitting function (\ref{eq:solve-power}) of the  Kaplan-Meier estimators.}
\end{table}
\smallskip

To test the validity of the model, as in (III), we then carried out simulations in Matlab of the stochastic urn transfer model using the values of $\lambda$ and $\rho$ shown in Table~\ref{table:lsq-km}. We chose the value $p=0.9$ for all simulations, after running some sample simulations with other values of $p$. The value of $p$ is not critical since, as can be seen from (\ref{eq:expand}), $p$ is merely a scaling factor.
The simulations were run for $114$ steps, one for each month, for each of the categories, and these were repeated $10^5$ times. For each category, we then calculated the average value of $F(i,t)$ for $i=1,2,\ldots,t$, over the $10^5$ runs. The results of nonlinear least-squares curve fitting to these average values are shown in Table~\ref{table:lsq-agg}. Comparing Table~\ref{table:lsq-agg} with Table~\ref{table:lsq-km} shows all the values of $\lambda$ and $\rho$ to be very close.

\smallskip

To check the closeness between the Kaplan-Meier estimates, the product data and the averaged simulation data, as in (IV), we performed three Kolmogorov-Smirnov 2-sample 2-tailed tests, as described in Section 6.6.4 of \cite{SIEG88}. Assuming the null hypothesis to be that the Kaplan-Meier estimates, the product data and the averaged simulation data all come from the same population distribution, the critical value at significance level $\alpha=0.05$ is $0.1801$ for a sample of $114$ (number of months). The $D$ values for the three pairwise tests are shown in Table~\ref{table:ks-test}.
It can be seen that, in all cases, the values of the test statistic $D$ are smaller than the critical value. Hence, we cannot reject the null hypothesis at significance level $\alpha=0.05$. The values in the Sim-Prod column show that the distributions of the product data and averaged simulation data are extremely close, which is unsurprising since these are both derived directly from our model. We observe that the values in the other two columns are generally smaller for categories with a larger number of queries. We also note that, even at significance level $\alpha=0.10$, where the critical value is $0.1616$, the null hypothesis cannot be rejected.

\begin{table}[ht]
\begin{center}
\begin{tabular}{|l|c|c|c|}\hline
Data set       & $\lambda$ & $\rho$  & $R^2$\\ \hline \hline
Business       & 1.5572    & -0.8321 & 0.9999 \\ \hline
Entertainment  & 1.7908    & -0.7455 & 0.9999 \\ \hline
Science        & 0.7585    & -0.8320 & 0.9999 \\ \hline
Shopping       & 0.4755    & -0.7615 & 0.9999\\ \hline
Sports         & 3.1998    & -0.8666 & 0.9999 \\ \hline
Travel         & 3.4849    & -0.9286 & 0.9999 \\ \hline
All Categories & 4.5555    & -0.9139 & 0.9999 \\ \hline
\end{tabular}
\end{center}
\caption{\label{table:lsq-agg} Nonlinear least-squares regression with fitting function (\ref{eq:solve-power}) of the simulated data, using $\lambda$ and $\rho$ from Table~\ref{table:lsq-km}.}
\end{table}
\smallskip

\begin{table}[ht]
\begin{center}
\begin{tabular}{|l|c|c|c|}\hline
Data set       & KM-Sim & KM-Prod & Sim-Prod \\ \hline \hline
Business       & 0.0567 & 0.0561  & 0.0037 \\ \hline
Entertainment  & 0.0189 & 0.0193  & 0.0030 \\ \hline
Science        & 0.1082 & 0.1039  & 0.0046 \\ \hline
Shopping       & 0.0756 & 0.0734  & 0.0052 \\ \hline
Sports         & 0.0268 & 0.0246  & 0.0044 \\ \hline
Travel         & 0.0741 & 0.0724  & 0.0055 \\ \hline
All Categories & 0.0250 & 0.0219  & 0.0031 \\ \hline
\end{tabular}
\end{center}
\caption{\label{table:ks-test} The $D$ values for a 2-sample, 2-tailed Kolmogorov-Smirnov tests.}
\end{table}
\smallskip

In order to fit a Weibull distribution to the averaged simulation data, following (V), we first adjusted the data as on the right-hand side of (\ref{eq:weibullk}), in order to incorporate the first two correction terms in the Euler-Maclaurin summation formula. The value of $k$ was chosen so that the error due to using the approximation $\ln(1+x) \approx x$ is small. We then calculated the right-hand side of (\ref{eq:weibullk}) for each value of $i$ from $k$ to $t$, using the values of $F(i+1,t+1)$ and $F(k,t+1)$ from the simulation, and the values of $\lambda$ and $\rho$ from Table~\ref{table:lsq-agg}. We then applied the exponential function to these values from (\ref{eq:weibullk}) for each $i$, and used non-linear regression to fit the Weibull distribution in (\ref{eq:weibull}).
We chose $k=10$ after inspecting the values of $\lambda$ and $\rho$ for various values of $k$ from 1 to 20, and comparing these to the corresponding values for the averaged simulation data in Table~\ref{table:lsq-agg}. The results for $k=10$ are shown in the rows of Table~\ref{table:agg-weibull}; the $R^2$ values indicate an almost perfect fit in all cases. It can be seen that the values of $\lambda$ and $\rho$ shown in Table~\ref{table:agg-weibull} for the Weibull fit to the adjusted simulation data closely match those shown in Table~\ref{table:lsq-agg} for the averaged simulation data. All the $\rho$ values are between $0$ and $-1$, as expected.

\begin{table}[ht]
\begin{center}
\begin{tabular}{|l|c|c|c|}\hline
Data set       & $\lambda$ & $\rho$  & $R^2$\\ \hline \hline
Business       & 1.5232    & -0.8264 & 0.9991 \\ \hline
Entertainment  & 1.7051    & -0.7320 & 0.9991 \\ \hline
Science        & 0.7419    & -0.8268 & 0.9992 \\ \hline
Shopping       & 0.4589    & -0.7538 & 0.9995 \\ \hline
Sports         & 3.1203    & -0.8608 & 0.9992 \\ \hline
Travel         & 3.4555    & -0.9262 & 0.9989 \\ \hline
All Categories & 4.4918    & -0.9105 & 0.9989 \\ \hline
 \end{tabular}
\end{center}
\caption{\label{table:agg-weibull} Nonlinear least-squares regression to a Weibull for the adjusted simulation data with $k=10$.}
\end{table}
\smallskip

\section{Concluding Remarks}
\label{sec:conc}

We have proposed a stochastic evolutionary urn model for survival analysis applications in the context of human dynamics.
In our model, actors (represented by balls) remain in the system and survive until they die (i.e. are discarded) according to a specified mortality distribution, which may take a general form. A solution to the equations describing the model was obtained in the form of a product (\ref{eq:expand}).
We then obtained a continuous approximation to the solution  (\ref{eq:renewal}) via the renewal equation from age-structured population dynamics.
This provides a continuous analogue to our discrete stochastic urn-based model.
Power-law mortality, which in the continuous case gives rise to the Weibull distribution (\ref{eq:weibull}),
was used to model the survival of popular search engine queries.
This could also be used to analyse the survival of Wikipedia editors \cite{ZHAN12}, as well as other data sets relating to human behaviour.

\smallskip

Generative models, such as the one we have presented, have the potential to explain observed social phenomena and, more specifically, social mechanisms and the emergence of collective behaviour, as discussed in the introduction. Moreover, they allow us to gain insight into the underlying processes and may also be useful for prediction purposes \cite{HEND01}. In this context, extending the survival model, as in the Cox proportional hazard model \cite{KLEI12}, to allow the inclusion of features (known as {\em risk factors}) could give rise to more accurate predictions \cite{LEE12}.

\section*{Acknowledgements}
We would like to thank Suneel Kingrani who collected the Google Trends data and computed the Kaplan-Meier estimates for the data set.
We also thank the referees for their constructive comments, which helped us to improve the paper.

\newcommand{\etalchar}[1]{$^{#1}$}


\begin{thebibliography}{CGW{\etalchar{+}}08}

\bibitem[AB02]{ALBE01}
R.~Albert and A.-L. Barab\'asi.
\newblock Statistical mechanics of complex networks.
\newblock {\em Reviews of Modern Physics}, 74:47--97, 2002.

\bibitem[Apo99]{APOS99}
T.M. Apostol.
\newblock An elementary view of {E}uler's summation formula.
\newblock {\em American Mathematical Monthly}, 106:409--418, 1999.

\bibitem[Bar05]{BARA05}
A.-L. Barab\'asi.
\newblock The origin of bursts and heavy tails in human dynamics.
\newblock {\em Nature}, 435:207--211, 2005.

\bibitem[Bar07]{BARA07}
A.-L. Barab\'asi.
\newblock The architecture of complexity: {F}rom network strucutre to human
  dynamics.
\newblock {\em IEEE Control Systems Magazine}, 27:33--42, 2007.

\bibitem[BG03]{BRAC03}
C.~Bracquemond and O.~Gaudoin.
\newblock A survey on discrete lifetime distributions.
\newblock {\em International Journal of Reliability, Quality and Safety
  Engineering}, 10:69--98, 2003.

\bibitem[BSV07]{BORN07a}
S.~B\"{o}rner, S.~Sanyal, and A.~Vespignani.
\newblock Network science.
\newblock {\em Annual Review of Information Science \& Technology (ARIST)},
  41:537--607, 2007.

\bibitem[CFL09]{CAST09}
C.~Castellano, S.~Fortunato, and V.~Loreto.
\newblock Statistical physics of social dynamics.
\newblock {\em Reviews of Modern Physics}, 81:591--646, 2009.

\bibitem[CGW{\etalchar{+}}08]{CAND08}
J.~Candia, M.C. Gonz\'alez, P.~Wang, T.~Schoenhar, G.~Madey, and A.-L.
  Barab\'asi.
\newblock Uncovering individual and collective human dynamics from mobile phone
  records.
\newblock {\em Journal of Physics A: Mathematical and Theoretical}, 41:224015,
  11pp, 2008.

\bibitem[Cha94]{CHAR94}
B.~Charlesworth.
\newblock {\em Evolution in age-structured populations}.
\newblock Cambridge Studies in Mathematical Biology: 13. Cambridge University
  Press, Cambridge, U.K., 2nd edition, 1994.

\bibitem[CM06]{CHES06}
A.G. Chessa and J.M.J. Murre.
\newblock Modelling memory processes and internet response times: {W}eibull or
  power-law?
\newblock {\em Physica A}, 366:539--551, 2006.

\bibitem[Col95]{COLO95}
H.~Colonius.
\newblock The instance theory of automaticity: {W}hy the {W}eibull?
\newblock {\em Psychological Review}, 102:744--750, 1995.

\bibitem[EB14]{EAGL14}
M.~Eagle and T.~Barnes.
\newblock Survival analysis on duration data in intelligent tutors.
\newblock In {\em Proceedings of International Conference on Intelligent
  Tutoring Systems (ITS)}, pages 178--187, Honolulu, HI, 2014.

\bibitem[FL00]{FLEM00}
T.R. Fleming and D.Y. Lin.
\newblock Survival analysis in clinical trials: {P}ast developments and future
  directions.
\newblock {\em Biometrics}, 56:971--983, 2000.

\bibitem[FLL07]{FENN05}
T.~Fenner, M.~Levene, and G.~Loizou.
\newblock A model for collaboration networks giving rise to a power-law
  distribution with an exponential cutoff.
\newblock {\em Social Networks}, 29:70--80, 2007.

\bibitem[FLL12]{FENN12}
T.~Fenner, M.~Levene, and G.~Loizou.
\newblock A discrete evolutionary model for chess players’ ratings.
\newblock {\em IEEE Transactions on Computational Intelligence and AI in
  Games}, 4:84--93, 2012.

\bibitem[FLL13]{FENN13}
T.~Fenner, M.~Levene, and G.~Loizou.
\newblock A bi-logistic growth model for conference registration with an early
  bird deadline.
\newblock {\em Central European Journal of Physics}, 11:904--909, 2013.

\bibitem[FLL14]{FENN14}
T.~Fenner, M.~Levene, and G.~Loizou.
\newblock A stochastic evolutionary model for survival dynamics.
\newblock {\em Physica A}, 410:595--600, 2014.

\bibitem[Gal08]{GALA08}
S.~Galam.
\newblock Sociophysics: {A} review of {G}alam models.
\newblock {\em Journal of Modern Physics C}, 19:409--440, 2008.

\bibitem[GG01]{GAVR01}
L.A. Gavrilov and N.S. Gavrilova.
\newblock The reliability theory of aging and longevity.
\newblock {\em Journal of Theoretical Biology}, 213:527--545, 2001.

\bibitem[HJS01]{HEND01}
R.~Henderson, M.~Jones, and J.~Stare.
\newblock Accuracy of point predictions in survival analysis.
\newblock {\em Statistics in Medicine}, 20:3083--3096, 2001.

\bibitem[HS98]{HEDS98}
P.~Hedstr\"{o}m and R.~Swedberg.
\newblock Social mechanisms: {A}n introductory essay.
\newblock In P.~Hedstr\"{o}m and R.~Swedberg, editors, {\em Social Mechanisms:
  An Analytical Approach to Social Theory}, pages 1--31. Cambridge University
  Press, Cambridge, U.K., 1998.

\bibitem[KK12]{KLEI12}
D.G. Kleinbaum and M.~Klein.
\newblock {\em Survival Analysis, A Self-Learning Text}.
\newblock Springer Science+Business Media, LLC, New York, NY, 3rd edition,
  2012.

\bibitem[KM58]{KAPL58}
E.L. Kaplan and P.~Meier.
\newblock Nonparametric estimation from incomplete observations.
\newblock {\em Journal of the American Statistical Association}, 53:457--481,
  1958.

\bibitem[LAE{\etalchar{+}}12]{LOVE12}
E.~Lovecchio, P.~Allegrini, E.Geneston, B.J. West, and P.~Grigolini.
\newblock From self-organized to extended criticality.
\newblock {\em Frontiers in Physiology}, 3:Article 98, 9pp, 2012.

\bibitem[Lam01]{LAMP01}
V.~Lampret.
\newblock The {Euler-Maclaurin} and {Taylor} formulas: {T}win, elementary
  derivations.
\newblock {\em Mathematics Magazine}, 74:109--122, 2001.

\bibitem[Lax06]{LAX06}
P.D. Lax.
\newblock {\em Hyperbolic Partial Differential Equations}.
\newblock Courant Lecture Notes. American Mathematical Society, Providence, RI,
  2006.

\bibitem[LB08]{LI08}
J.~Li and F.~Brauer.
\newblock Continuous-time age-structured models in population dynamics and
  epidemiology.
\newblock In F.~Brauer, P.~{van den Driessche}, and J.~Wu, editors, {\em
  Mathematical Epidemiology}, Lecture Notes in Mathematics, Mathematical
  Biosciences Subseries, chapter~9, pages 205--227. Springer-Verlag, Berlin,
  2008.

\bibitem[LMS12]{LEE12}
J.G. Lee, S.~Moon, and K.~Salamatian.
\newblock Modeling and predicting the popularity of online contents with {C}ox
  proportional hazard regression model.
\newblock {\em Neurocomputing}, 76:134--145, 2012.

\bibitem[LS98]{LAHE98}
J.~Laherr{\`e}re and D.~Sornette.
\newblock Stretched exponential distributions in nature and economy: “fat
  tails” with characteristic scales.
\newblock {\em European Physical Journal B}, 2:525--539, 1998.

\bibitem[MAAJ13]{MATH13}
J.~Mathiesen, L.~Angheluta, P.T.H. Ahlgren, and M.H. Jensen.
\newblock Excitable human dynamics driven by extrinsic events in massive
  communities.
\newblock {\em Proceedings of the National Academy of Sciences of the United
  States of America}, 110:17259--17262, October 2013.

\bibitem[Mot95]{MOTU95}
H.~Motulsky.
\newblock {\em Intuitive Biostatistics}.
\newblock Oxford University Press, Oxford, 1995.

\bibitem[MSR13]{MITN13}
A.~Mitnitski, X.~Song, and K.~Rockwood.
\newblock Assessing biological aging: the origin of deficit accumulation.
\newblock {\em Biogerontology}, 14:709--717, 2013.

\bibitem[New05]{NEWM05}
M.E.J. Newman.
\newblock Power laws, {P}areto distributions and {Z}ipf's law.
\newblock {\em Contemporary Physics}, 46:323--351, 2005.

\bibitem[OK12]{OCON12}
P.D.T. {O'Connor} and A.~Kleyner.
\newblock {\em Practical Reliability Engineering}.
\newblock Wiley Series in Telecommunications. John Wiley {\&} Sons, Chichester,
  5th edition, 2012.

\bibitem[PR91]{PILA91}
M.~Pilant and W.~Rundell.
\newblock Determining a coefficient in a first-order hyperbolic equation.
\newblock {\em SIAM Journal on Applied Mathematics}, 51:494--506, 1991.

\bibitem[PS10]{PAN10}
R.K. Pan and S.~Sinha.
\newblock The statistical laws of popularity: universal properties of the
  box-office dynamics of motion pictures.
\newblock {\em New Journal of Physics}, 12:115004 (23pp), 2010.

\bibitem[Rin09]{RINN09}
H.~Rinne.
\newblock {\em The Weibull Distribution: A Handbook}.
\newblock CRC Press, Boca Raton, Fl., 2009.

\bibitem[RNP{\etalchar{+}}10]{RICH10}
J.T. Rich, J.G. Neely, R.C. Paniello, C.C.J. Voelker, B.~Nussenbaum, and E.W.
  Wang.
\newblock A practical guide to understanding {Kaplan-Meier} curves.
\newblock {\em Otolaryngology–Head and Neck Surgery}, 143:331--336, 2010.

\bibitem[Ros96]{ROSS96}
S.M. Ross.
\newblock {\em Stochastic Processes}.
\newblock John Wiley {\&} Sons, New York, NY, 2nd edition, 1996.

\bibitem[SBCG13]{SCHN13}
C.M. Schneider, V.~Belik, T.~Couronn\'e, and M.C. Gonz\'alez.
\newblock Unravelling daily human mobility motifs.
\newblock {\em Journal of the Royal Society Interface}, 10:20130246, 2013.

\bibitem[SC88]{SIEG88}
Siegel S and N.J. {Castellan Jr.}
\newblock {\em Nonparametric Statistics for the Behavioral Sciences}.
\newblock McGraw-Hill, New York, NY, 2nd edition, 1988.

\bibitem[Sch98]{SCHE98a}
T.C. Schelling.
\newblock Social mechanisms and social dynamics.
\newblock In P.~Hedstr\"{o}m and R.~Swedberg, editors, {\em Social Mechanisms:
  An Analytical Approach to Social Theory}, pages 32--44. Cambridge University
  Press, Cambridge, U.K., 1998.

\bibitem[Sim55]{SIMO55}
H.A. Simon.
\newblock On a class of skew distribution functions.
\newblock {\em Biometrika}, 42:425--440, 1955.

\bibitem[TW02]{TSOU02}
A.~Tsoularis and J.~Wallace.
\newblock Analysis of logistic growth models.
\newblock {\em Mathematical Biosciences}, 179:21--55, 2002.

\bibitem[VAFL10]{MELO10}
P.O.S. {Vaz De Melo}, L.~Akoglu, C.~Faloutsos, and A.A.P. Loureiro.
\newblock Surprising patterns for the call duration distribution of mobile
  phone users.
\newblock In {\em Proceedings of European Conference on Machine Learning and
  Principles and Practice of Knowledge Discovery in Databases (ECML PKDD)},
  pages 354--369, Barcelona, 2010.

\bibitem[ZPL12]{ZHAN12}
D.~Zhang, K.~Prior, and M.~Levene.
\newblock How long do wikipedia editors keep active?
\newblock In {\em Proceedings of the 8th International Symposium on Wikis and
  Open Collaboration (WikiSym)}, Linz, Austria, 2012.

\end{thebibliography}
\end{document}